# Two-stage catalytic hydrotreatment of highly nitrogenous biocrude from continuous hydrothermal liquefaction: A rational design of the stabilization stage

Muhammad Salman Haider, Daniele Castello, Lasse Aistrup Rosendahl

*Department of Energy Technology, Aalborg University, Pontoppidanstræde 111, 9220 Aalborg Øst, Denmark*

# ACCEPTED MANUSCRIPT

# Two-stage catalytic hydrotreatment of highly nitrogenous biocrude from continuous hydrothermal liquefaction: A rational design of the stabilization stage

Muhammad Salman Haider, Daniele Castello[*], Lasse Aistrup Rosendahl

*Department of Energy Technology, Aalborg University, Pontoppidanstræde 111, 9220 Aalborg Øst, Denmark*

**Abstract:** Effective catalytic hydrotreatment of highly nitrogenous biocrudes derived from the hydrothermal liquefaction (HTL) of primary sewage sludge and microalga *Spirulina* biomass was explored. A critical issue is the lack of thermal stability of raw HTL biocrudes at the severe conditions (~400 °C) required for hydrodenitrogenation. This fact suggests the need for a two-stage approach, involving a first low-temperature stabilization stage followed by another one operated at higher temperature. In this study, DSC was successfully used to indicate the thermal stability of both biocrudes. During hydrotreating, it was observed that complete deoxygenation was already achieved in the first stage at 350 °C, with limited coke formation. Moreover, after second stage up to 92% denitrogenation associated with the higher hydrogen consumption (39.9 g kg$^{-1}$ for *Spirulina* and 36.9 g kg$^{-1}$ for sewage sludge) was obtained for both biocrudes. Consequently, comparable oil yields but significantly less coke yields were recorded during two-stage upgrading (1.0% for *Spirulina* and 0.7% for sewage sludge), compared to direct processing at 400 °C (9.1% for *Spirulina* and 3.4% for sewage sludge). In addition, the properties of the upgraded oils were enhanced by increasing the temperature in the first stage (310 °C, 330 °C and 350 °C respectively). Finally, the results indicated that remarkable drop-in fuel properties were obtained, with respect to heteroatom (O and N) removal, HHV, and H/C ratio during the two-stage hydrotreatment. Two-stage hydrotreating is therefore proposed as a successful approach for the upgrading of HTL biocrudes with high nitrogen content.

**Keywords:** Hydrothermal liquefaction (HTL); *Spirulina* biocrude; Sewage sludge biocrude; Coke formation; Catalytic hydrotreating; Drop-in biofuels.

---

[*] Corresponding author. E-mail: dac@et.aau.dk

# 1. Introduction

Hydrothermal liquefaction (HTL) is an effective valorization technique, capable of efficiently processing wet/dry biomass with no lipid content restriction, ranging from lignocellulosic over algae to organic wastes. The output from HTL is an energetically dense intermediate product, referred to as biocrude [1,2]. However, regardless of its origin, HTL biocrude usually presents a high heteroatom content (primarily oxygen and nitrogen). In its raw form it is not compatible with most uses in the transport sector, therefore it needs significant upgrading [3].

Upgrading by catalytic hydrotreating is the most promising pathway, but at the same time it represents a serious scientific and engineering challenge for the conversion of HTL biocrude into hydrocarbons [4–6]. During hydrotreating, heteroatoms in the biocrude are removed by using hydrogen in the presence of heterogeneous catalysts at elevated temperatures (300-450 °C) and pressures (3-17 MPa) [7].

The presence of considerable amounts of oxygenated compounds in HTL biocrudes [3,8] could make them unstable under severe conditions [4,9], which means they could be susceptible to polymerization/condensation or cracking reactions as observed in pyrolysis bio-oils [10–12]. Hydrotreatment removes oxygen through dehydration, decarboxylation and decarbonylation reactions and eliminates nitrogen mainly in the form of ammonia gas [13,14]. At severe conditions, coking appears to be much faster, suggesting competition between hydrogenation/hydrodeoxygenation/hydrodenitrogenation and coking reactions. This could lead to rapid catalyst deactivation, low hydrocarbon yields, formation of undesirable products (e.g. coke) and, eventually, reactor plugging [9]. Moreover, nitrogen-containing compounds in the biocrude are often encountered, among the others, in the form of heterocyclic compounds [3,15]. Apart from increasing the chemical complexity, these compounds also lead to a higher $H_2$ consumption, as they are recalcitrant to denitrogenation. Indeed, the reaction mechanism involves high temperature (≥400 °C) and high $H_2$ pressure for the saturation of aromatic rings prior to nitrogen removal [16–18]. N-compounds in HTL biocrude can be therefore difficult to remove, thus requiring more severe operating conditions in terms of temperature and partial $H_2$ pressure [19,20]. As HTL biocrudes are generally more stable than the bio-oils from pyrolysis due to their lower oxygen content [21], no real attention has been devoted to the issue of coke formation at higher temperatures (~400 °C) during their upgrading through hydrotreatment.

In the open literature, no study with a focus on the thermal stability of HTL biocrudes and its adverse impacts during hydrotreatment at high temperatures is documented. However, a number of studies have been performed on the hydrotreatment of HTL biocrude from algae in the presence of different catalysts [22–25], where the focus was on achieving a high degree of heteroatoms removal. Zhao et al. [26,27] investigated the catalytic upgrading of algae biocrude in two-stage batch experiments, and reported almost complete removal of N (99.5%) in the presence of a commercial NiMo/γ-$Al_2O_3$ catalyst. Cole et al. [28] successfully carried out two-stage upgrading for macroalgae biocrude and reported it as the only viable way to effectively achieve drop-in fuels. Haghighat et al. [13] also studied the two-stage catalytic upgrading of HTL biocrude from forestry residue in a continuous fixed bed reactor, and documented 83.5% removal of oxygen and decrease of residue fraction (boiling point >550 °C) from 35% to 5% in weight by using commercial CoMo/γ-$Al_2O_3$ and NiMo/γ-$Al_2O_3$ catalyst in first and second stage respectively. Researchers from the Pacific Northwest National Laboratories (PNNL) hydrotreated different algal biocrudes in a continuous trickle bed reactor by using two different temperature zones (i.e. 120-170 °C and 405 °C) in a single reactor under 13.6 MPa of $H_2$ pressure [20]. Furthermore, Albrecht et al. [14] and Marrone et al. [29] successfully evaluated the continuous hydrotreatment of algae and primary sewage sludge biocrude in a single temperature zone of nominally 400 °C with the $H_2$ consumption of 51 g kg$^{-1}$ and 44 g kg$^{-1}$. In all cases, they obtained almost complete removal of nitrogen (~99%), with a residual amount of oxygen (~1%).

As far as coke formation and reactor plugging is concerned, Jensen [9] reported high exothermicity, significant pressure drop and catalyst deactivation due to severe coking propensity within the first ten hours in the first reactor, during the two-stage continuous hydrotreatment of lignocellulosic HTL biocrude at 360 °C and 10 MPa $H_2$ pressure. However, they achieved smooth hydroprocessing for 330 hours while operating between 310 °C to 330 °C. Bai et al. [30] also reported the formation of coke (~15-29Figure 3 %), during the hydrotreatment of algae biocrude by using 14 different commercial and noble metal catalysts. Recently, Castello et al. [4] reported the coke yields during the catalytic upgrading of HTL biocrude from algae and primary sewage sludge as a function of temperature and pressure. Moreover, Biller et al. [22] documented low yields (41-69%) of upgraded biocrude and observed coke after the hydroprocessing of microalgae at 405 °C and documented 55-60% denitrogenation. It is therefore evident from the literature that direct hydroprocessing of HTL biocrudes at high temperatures results in significant coking, and this could potentially lead to reactor plugging. Therefore, a rational experimental approach is followed to address this core issue.

The practical pathway adopted in this paper involves the sequential hydrotreatment of nitrogen-rich HTL biocrudes (i.e. *Spirulina* algae and primary sewage sludge), in which both biocrudes are first stabilized at lower temperatures (310-350 °C) by removing the oxygen-containing compounds via hydrodeoxygenation reaction. Subsequently, this stabilized product is further processed in a second stage at higher temperature (400 °C) with minimal coke production as compared to direct hydrotreatment at high temperature (400 °C). The specific objective is to bring forward the understanding of thermal stability and tendency to coke formation during the direct hydroprocessing of HTL biocrudes at severe conditions (i.e. 400 °C). In present study, thermogravimetric analysis coupled with differential scanning calorimetry (DSC) was used to indicate the thermal stability of both HTL biocrudes at 350 °C and 400 °C. This information appeared to be significant in relation to coke formation during the direct hydroprocessing of HTL biocrudes at 350 °C and 400 °C. Additionally, by changing temperatures in the first stage, its effect on the coke production and oil yield with better drop-in fuel properties were also studied. Results show that the formation of coke during direct hydrotreatment at high temperatures is a real challenge, but it can be overcome by using multistage hydrotreatment.

## 2. Materials and Methods

### 2.1. Materials

In the present study, the HTL biocrudes from microalgae *Spirulina* (Otog, Ordos, Inner Mongolia, China (Inner Mongolia Rejuve Biotech Co. Ltd.)) and primary sewage sludge (wastewater treatment plant at Viborg, Denmark (Energi Viborg A/S)) were obtained from Aarhus University, Denmark. These biocrudes were produced at sub-critical conditions (22 MPa, 350 °C) in a continuous HTL pilot plant. Details related to the properties and the production of these two biocrudes can be found in Ref. [31].

Received biocrudes were first filtered, in order to reduce their ash content. This operation was utilized for both reducing the viscosity of the feed and removing minerals that could negatively affect catalyst activity.

Biocrudes were first diluted with acetone and then vacuum filtered over VWR filter paper (5-13 μm). Solvent was finally removed by vacuum distillation at 60 °C and 50 kPa, with a rotary evaporator (Büchi AG). This procedure was especially useful in the case of sewage sludge biocrude, with a dramatic ash content reduction from 28.6% down to 0.06%. A minor removal was recorded for *Spirulina*: from 0.17% to 0.14%. Ash content was determined according to ASTM D482 [32], through incineration at 775 °C in a Protherm electric muffle furnace.

A pre-sulfided trilobe NiMo/γ-$Al_2O_3$ commercial hydrotreating catalyst (Criterion Centera™ DN-3630) in the form of 1 mm trilobe extrudates was provided by Shell refinery in Fredericia, Denmark.

## 2.2. Hydrotreating Experiments

In order to understand the thermal stability of both biocrudes and its impact on hydrotreatment, an experimental approach based on two-stage batch hydrotreatment was adopted. This approach consists in performing a mild hydrotreatment (310 °C, 330 °C and 350 °C) batch experiment (first stage), collecting the produced upgraded oil (excluding the water-phase) and using it for a subsequent severe hydrotreatment experiment at 400 °C (second stage). The overall results obtained from two-stage processing were compared with those of single-stage direct hydrotreatment, conducted at severe conditions (400 °C). A detailed list of test runs is reported in Table 1, where, both reaction time (4 h) and initial $H_2$ pressure (8 MPa) are kept constant, based on the findings from previous work by Haider et al. [19].

TABLE 1

Hydroprocessing reactions were carried out in 25 $cm^3$ Swagelok micro-batch stainless steel tube reactors. In order to assure reproducibility and comparability of results, all experiments were performed as duplicates by using two identical reactors. Mean values of the results were reported for all experiments. However, the absolute difference between mean value and measured experimental values was reported as error. Errors were used to show the accuracy of the experimental results. In each experiment (first or second stage) each reactor was loaded with 4 g of biocrude and 2 g of pre-activated NiMo/γ-$Al_2O_3$ catalyst. In order to enhance mixing during hydroprocessing, three stainless steel spheres (4 mm dia.) were also loaded in each reactor. The reactors were then sealed and purged with $N_2$ and $H_2$ twice at 10 MPa, leak tested and pressurized with the desired initial $H_2$ pressure. A fluidized sand-bath SBL-2D (Techne, UK) and an agitation device with a frequency of 450 min$^{-1}$ were used to obtain desired temperatures and efficient mixing during whole reaction time. Wika A-10 pressure transducers coupled with a LabView™ program were used to continuously record pressure during the reaction. All reactions were carried out for 4 h with 4 g of feed and 2 g of catalyst in order to mimic a continuous hydrotreating operation with a weight hourly space velocity (WHSV) of 0.5 h$^{-1}$ [4,14].

After the desired reaction time, reactors were quenched in a water bath and then dried with compressed air prior to gas venting. Gas yield was estimated by weighing the reactor before and after gas venting [30]. Moreover, the liquid products (i.e. oil and water phases) were carefully collected and weighed after passing them through a metallic mesh in order to separate the catalyst and the carbonaceous solids (which will be indicated as "coke") formed during the reaction. Oil and water phases were then separated by centrifugation (Sigma 6-16 HS centrifuge, 2153 RCF) for 5 min. Furthermore, the interior of the reactor and the catalyst were washed with ~100 cm$^3$ of dichloromethane and the resulting mixture was vacuum filtered on paper (VWR). Catalyst and solids were recovered on the filter and then dried overnight at 120 °C, in order to determine the amount of carbonaceous solids by weight difference. However, the amount of coke was calculated by subtracting the inorganics from solids. Dichloromethane was separated from the filtered solution by distillation in a rotary evaporator (Büchi) at 60 °C and 90 kPa. The overall oil yield was thus estimated as the sum of the oil from direct collection and the one recovered after filtration and dichloromethane evaporation. However, analyses were only carried out on the oil samples from direct collection, i.e. those which did not undergo filtration and solvent extraction, in order to minimize any possible deviation from their original composition.

## 2.3. Characterization and Analytical Techniques

Initial feeds, as well as the reaction products, were thoroughly characterized and compared by using several analytical instruments. At first, the thermal stability of both HTL derived biocrudes was investigated by using thermogravimetric analysis combined with differential scanning calorimetry (TGA-DSC). Data points were collected with a SDT 650 simultaneous DSC/TGA (TA Instruments Discovery) using a sample size of

~15 mg. Samples were placed in ceramic crucibles and heated at 5 K min$^{-1}$. Biocrude samples were analyzed by using isothermal runs at 350 °C and 400 °C for 230 min with a N$_2$ flow rate of 50 cm$^3$ min$^{-1}$, following the approach reported in ref. [33].

The gaseous products were analyzed by using a GC-2010 (Shimadzu Inc.) gas chromatograph with a barrier ionization discharge detector (GC-BID), and equipped with a Supelco 1006 PLOT column. After the hydrotreating reaction, the measured H$_2$ concentration in the gas along with the initial and final gas pressures in the reactor was used to estimate the overall H$_2$ consumption. The ideal gas law was utilized for this purpose. Moreover, it was assumed that 20 cm$^3$ of the reactor volume were occupied by the gas. It should be noted that the detection of hydrogen sulfide and ammonia gas (i.e. product of hydrodesulfurization and hydrodenitrogenation) in this study was left unaccounted due to the limitations of the measuring instrument. A PerkinElmer® 2400 Series II CHN-O Elemental Analyzer, following ASTM D5291 [34] with a detection limit of 0.01% was used to determine the elemental composition of raw and upgraded biocrudes in terms of C, H and N. Oxygen was calculated by difference. The amount of coke, i.e. carbonaceous solids found after reaction, was calculated by subtracting the solids from inorganic content associated with both biocrudes. Moreover, the sulfur (S) in raw and upgraded biocrudes was measured by using HORIBA SLFA-60 X-ray Fluorescence Analyzer with a detection limit of 0.005%, according to the standard ASTM D4294 [35].

Furthermore, the degree of deoxygenation (de-O) and denitrogenation (de-N) were calculated as follows:

$$\text{de-O} = \left(1 - \frac{O_{product}}{O_{biocrude}}\right) \cdot 100 \qquad (1)$$

$$\text{de-N} = \left(1 - \frac{N_{product}}{N_{biocrude}}\right) \cdot 100 \qquad (2)$$

Channiwala-Parikh correlation [36], based on the elemental mass fraction, was used to estimate the high heating value (HHV) of both raw and upgraded biocrudes:

$$\text{HHV (MJ kg}^{-1}) = 0.3491C + 1.1783H - 0.1034O - 0.0151N \qquad (3)$$

The boiling point distributions of both raw and upgraded products were determined by means of simulated distillation (Sim-Dis), according to the standard ASTM D7169 [37]. The apparatus used for this analysis was a GC-2010 (Shimadzu Inc.) gas chromatograph with a flame ionization detector (GC-FID), equipped with a Zebron ZB-1XT column (Phenomenex). The chemical composition of oil samples was analyzed by gas chromatography coupled with mass spectrometry (GC-MS). GC-MS (Thermo Scientific Trace 1300 ISQ QD - Single Quadrupole) was equipped with a HP-5MS column. Details on samples preparation and GC-MS method were described previously [4]. Before and after hydrotreating, different chemical families were observed by using GC-MS chromatogram. For this purpose, the largest 200 peaks (excluding solvent) in terms of chromatogram area were evaluated.

## 3. Results and Discussion

Hereby, the results of different hydrotreating experiments conducted on *Spirulina* algae (1.216±0.005% S) and primary sewage sludge (0.574±0.003% S) biocrudes are presented. At first, the thermal stability of HTL biocrudes is described. Subsequently, the relevance to two-stage hydrotreatment, in relation with oil yield, coke formation, heteroatom removal and hydrogen consumption is discussed in detail. Finally, the raw and upgraded biocrudes are compared in terms of chemical composition and boiling point distribution.

### 3.1. Thermal stability at high temperatures

Considerable amount of oxygen-containing compounds are present in untreated HTL biocrudes from *Spirulina* and sewage sludge. In both biocrudes oxygen is mainly present in the form of fatty acids, alcohols, phenols and carbonyl-containing compounds [3,4,8,15]. In bio-oils, carbonyl and phenol compounds are responsible for chemical and thermal instability, and they are quite known for coke formation [38–42]. Because of the high reactivity of these oxygen-containing compounds, heat-induced chemical changes in HTL biocrudes are investigated as described in literature related to bio-oils [33,43].

DSC was used to monitor the net heat absorbed or released from both biocrudes, where the former relates to endothermic reactions and the latter exhibits exothermic reaction. Therefore, DSC was used to examine the thermal stability of both HTL biocrudes. Figure 1 illustrates isothermal DSC plots of both biocrudes at 350 °C and 400 °C. These temperatures were selected based on observations reported by Biller et al. [22], Bai et al. [30] and Castello et al. [4], where they observed coke formation at 400 °C and reported higher yields of liquids at 350 °C.

At 350 °C, the isothermal DSC curve remains steady and no changes in the heat flow signals were observed within 230 min, indicating the absence of reactions. Thus, it was reasonable to conclude that both biocrudes are thermally stable at 350 °C. As opposite, changes in the heat flow were observed with both endothermic and exothermic peaks throughout the isothermal DSC curve at 400 °C. Consequently, both biocrudes are thermally unstable at these severe conditions and are susceptible to polymerization/condensation or cracking reactions. This thermal instability in both biocrudes might be caused, primarily, due to the presence of highly reactive oxygen-containing functional groups [8,15] and aromatics [44] (especially for *Spirulina*). The information about the thermal stability of HTL biocrudes under desired hydrotreating conditions, could significantly improve our understanding about possible polymerization of HTL biocrudes at these high temperature conditions. In literature, batch hydrotreatment of HTL biocrudes at severe reaction conditions (400 °C) produces significant amount of coke [4,30], which may be caused due to undesired reaction pathways associated with severe thermal conditions. The results from DSC suggest that direct thermal treatment at severe conditions is not an optimal route, while prior processing at lower temperature can possibly result in smoother operations.

FIGURE 1

## 3.2. Effect of temperature on oil yields vs. coke formation

To evaluate the effect of temperature on oil yield and coke formation, both biocrudes were hydrotreated in two stages and the results were compared with the single-stage experiment performed at 400 °C. Hydrotreatment strongly affected the physical appearance of the oil samples, as it can be seen in Figure 2. Indeed, the raw biocrude is a black and highly viscous liquid (Figure 2a). The first hydrotreating stage saturated the complex aromatic ring structures and removed most of the oxygen-containing compounds via hydrodeoxygenation, and lowered the average molecular mass. This causes its viscosity to decrease and its color to become lighter (Figure 2b). However, after the second stage, the appearance of the biocrude changes dramatically. This may be due to the removal of nitrogen-containing compounds via hydrodenitrogenation. Indeed, some of these compounds (e.g. amines and indoles, generally present in biocrudes [3,4]) are known as strong organic dyes [45,46]. Now, the upgraded biocrude is a non-viscous, light yellow liquid (Figure 2c). The appearance of the oil is thus strongly affected by the degree of heteroatoms removal and hydrogenation achieved in the reaction.

FIGURE 2

As shown in Table 2, the temperature of the first stage had a critical influence on the oil yields and coke formation. Hydrotreatment of both biocrudes at temperatures ensuring thermal stability (Figure 1) resulted in dramatic decreases in coke formation, as compared to the single-stage experiments at 400 °C. Indeed, at these conditions (single-stage experiment at 400 °C), lower oil yields and higher gas yields were encountered, which is due to the extensive coke formation due to polymerization reactions and higher extent of cracking.

At 310 °C, the least possible coke was formed for both biocrudes and it only slightly increased while operating at 330 °C and 350 °C. However, at 400 °C (single-stage) 9.1% coke yield was obtained in case of *Spirulina* and 3.4% of coke was collected for sewage sludge biocrude. This relevant fact can be described on the basis of chemical constituents present in both biocrudes. Sewage sludge biocrude mostly contains fatty acids, alcohols and esters, whereas *Spirulina* biocrude also shows a remarkable presence of phenols, carbonyl compounds and other oxygenated aromatics [3,4,15]. Thus, the higher aromaticity and the presence of highly reactive oxygen-containing compounds [8,9,15,44] in *Spirulina* biocrude make it more vulnerable towards coke formation, when exposed to severe conditions (400 °C) during hydrotreatment. Coke formation could also be favored by mass transfer limitations, which are likely to be present in batch catalytic hydrotreaters [12,22]. However, it must be pointed out that extensive coke formation during the hydrotreating of demineralized HTL biocrude was also observed by Jensen [9] in a continuous fixed-bed reactor, where much better mass transfer could be achieved. Therefore, although mass transfer can play some role, the intrinsic coking tendency of biocrude at high temperature seems to be the determinant feature.

Table 2 shows intriguing results when a two-stage approach is adopted for the hydrotreatment of both biocrudes. Under the conditions studied, coke yields in all experiments at second stage became negligible, ranging from 0.014% to 0.05%. Then, it is reasonable to assume that the undesired coke formation can be suppressed by reducing the concentration of reactive components via hydrogenation, and by selecting the temperatures in which HTL biocrudes are thermally stable. Hence, in an overall two-stage process, the total amount of coke produced will be represented by the sole quantity obtained in the first stage, whose temperature should be carefully selected.

During hydrotreatment, oxygen is removed via several parallel reactions including hydrodeoxygenation, decarboxylation and decarbonylation [4,13]. Water yields in Table 2 are directly related to the extent of oxygen removal through hydrodeoxygenation, and therefore the yields of water increase with the severity of reaction temperatures in first stage. For *Spirulina*, most of water was produced already at 310 °C, with small increase at 330 °C and 350 °C. However, sewage sludge biocrude required 350 °C to remove most of oxygen in the form of water. During single-stage (direct) hydrotreating of both biocrudes, water yields were comparable at 350 °C and 400 °C. This suggests that most oxygen is already removed at mild reaction conditions i.e. 350 °C, which makes the partial upgraded products more stable for further temperature treatment under severe conditions. Moreover, by comparing the oxygen content in biocrude feeds (Table 3) and the water yields after hydrotreating, it can be appreciated that hydrodeoxygenation is markedly the preferred pathway for oxygen rejection. These results are in line with the finding of Haghighat et al. [13].

TABLE 2

Figure 3 shows a graphical comparison of the overall oil and coke yields after each stage of treatment, based on the feed of the first stage. Comparable oil yields were obtained while operating with both two-stage hydrotreating and single-stage at 400 °C. Differently, two-stage hydrotreating greatly reduces coke formation compared to single-stage at 400 °C. From the perspective of oil yield vs. coke formation, the two-stage hydrotreating experiment with 310 °C in the stabilization stage was the best choice for *Spirulina*. As far as the hydrotreatment of sewage sludge biocrude is concerned, the two-stage experiment with 330 °C in the first stage was more effective. On the other hand, the choice of a higher temperature in the first stage can significantly improve heteroatoms removal, as it will be shown in Section 3.3.

FIGURE 3

### 3.3. Heteroatoms removal, $H_2$ consumption and carbon loss

Table 3 shows the results of the elemental analysis of raw and upgraded biocrude samples after the hydrotreating experiments. Heteroatoms removal appears to be the most challenging aspect of biocrude upgrading. For both biocrudes remarkable results were attained in the stabilization stage, where 100% de-O and more than 50% of de-N were reached at 350 °C. However, the single-stage treatment at 400 °C also showed complete deoxygenation, although with a lower H/C and HHV compared to the respective experiments at 350 °C. This implied that the direct exposure of both biocrudes at 400 °C was effective in the removal of O and N, but at the expense of coke production and increased aromaticity in the fuel quality. The subsequent hydrotreating step after the stabilization stage resulted in a higher de-N, with the removal of ~92% of nitrogen content in both biocrudes. For both biocrudes, it is observable that deoxygenation is relatively easier to achieve than denitrogenation, in good accordance with the literature [4,19,22]. Moreover, after two-stage hydrotreatment complete desulfurization was achieved, whereas, after single-stage hydrotreatment (i.e. 400 °C) 96% and 97% of sulfur was removed from *Spirulina* and sludge biocrude respectively.

TABLE 3

Measured values of $H_2$ consumption are shown in Figure 4, along with de-O and de-N efficiencies. As a general trend, it appears that $H_2$ consumption is directly related to heteroatoms removal. Overall $H_2$ consumption after second stage (i.e. first stage at 350 °C) is about 39.9 and 36.9 g kg$^{-1}$ for *Spirulina* and sludge respectively, which is significantly higher than the single-stage hydrotreatment at 400 °C (i.e. 21.3 and 19.9 g kg$^{-1}$ for *Spirulina* and sludge respectively). The second stage shows substantial amount of hydrogen consumption, which directly relates to the higher extent of denitrogenation but, at the same time, to the higher degree of hydrogenation [17,47], indicated by the increase in the H/C ratio. From Figure 4, it is also evident that $H_2$ consumption remains approximately constant during the single-stage hydrotreatment at 350 °C and 400 °C. The overall values of $H_2$ consumption obtained after two-stage batch hydrotreatment are comparable to those obtained by Albrecht et al. (51 g kg$^{-1}$) [14] and Marrone et al. (44 g kg$^{-1}$) [29] in a continuous device. It should be remarked that they observed almost complete removal of nitrogen (~99%), whereas, in present study, ~92% of nitrogen removal is achieved.

Furthermore, the effect of $H_2$ consumption (Figure 4) is quite significant and relates directly with the coke formation (Figure 3) when compared with the direct hydrotreatment at 400 °C. During direct hydrotreatment at 400 °C the higher coke yields are responsible for lower $H_2$ consumption, probably due to catalyst fouling.

FIGURE 4

TABLE 4

The carbon balances of all hydrotreating experiments including gases are reported in Table 4. It can be observed that direct hydrotreatment at severe conditions results in reduced yields, due to the higher extent of coke formation and heteroatom removal. However, in two-stage hydrotreating process the reduced yields are primarily associated with higher extent of heteroatom removal. In all cases, only minor carbon loss is associated with the decarboxylation ($CO_2$) and decarbonylation (CO) reaction. This confirms the results reported in Table 2 which indicate that most of the oxygen is removed in the form of water. Moreover, carbon loss is also associated with the inherent losses during handling and sample separation after the micro-batch hydrotreating experiment. More prominently, this effect can be noticed after two-stage hydrotreating experiment where the first stage was operated at 350 °C. Indeed, after second stage the share of volatile compounds in the upgraded oil greatly increased [24], as it can be observed in Figure 6.

The relation between denitrogenation and hydrogenation can be better understood by means of a modified van Krevelen plot in terms of H/C and N/C atomic ratios [19,48]. From Figure 5, it is clear that temperature has a significant effect on both nitrogen removal and the reduction of the aromatic content (especially in *Spirulina*). In general, higher denitrogenation is associated with a higher degree of saturation [17,47], which corresponds to a higher H/C ratio. This can be evidently noticed for all second-stage experiments, with a remarkable effect of the first stage temperature. In particular, two-stage tests involving stabilization at 350 °C showed a remarkable increase of H/C ratio (i.e. 2.07 and 2.15 for *Spirulina* and sludge respectively) in the second stage. This indicates a strong reduction in the aromatic content in the upgraded oil but, at the same time, it indicates that a considerable share of the $H_2$ used in the process is utilized for C-H bonds saturation, rather than for heteroatoms removal. The high values of H/C reveal that the upgraded biocrudes are in paraffinic range [49]. However, in terms of upgraded oil quality, the two-stage hydrotreatment with 350 °C in the stabilization stage appears to be preferable for both biocrudes, as it provides maximum HHV, highest H/C and lowest N/C atomic ratio for both biocrudes.

FIGURE 5

## 3.4. Analysis of biocrude and upgraded products

Simulated distillation (Sim-Dis) was used to obtain the boiling point distribution of both raw and upgraded biocrudes. Figure 6 shows the boiling point distribution of raw and upgraded biocrude samples. *Spirulina* biocrude looks lighter than the sewage sludge one, as it is testified by their measured recoveries, i.e. 72% and 63% respectively. This fact becomes even more evident if the temperature of 320 °C is considered, corresponding to products up to the light diesel range. At this temperature, sewage sludge biocrude exhibits lower recovery (6%) at 320 °C, compared to *Spirulina* (21%). This aspects reveals that sewage sludge biocrude contains higher molecular weight compounds.

Processing at high temperatures shifts the boiling point distribution of both biocrudes towards lower boiling points and the overall recoveries significantly increase. To a large extent, the change in the boiling point distribution is observed during the first stage of hydrotreatment, where stabilization takes place. As it was observed in Sections 3.2 and 3.3, almost complete deoxygenation is reported during the stabilization stage. Consequently, it can be seen that deoxygenation represents the most prominent factor dominating the boiling point distribution.

Boiling point distribution is strongly affected by temperature. As it can be expected, higher processing temperatures result in a higher extent of cracking, leading to the formation of compounds with a lower boiling point. As it can be visualized in Figure 6, boiling point distributions show a dramatic increase during the first stage of hydrotreatment. This is due to the combined effect of the cracking of heavy compounds into smaller molecules (which is obtained as a consequence of heteroatoms removal) and to the reduction of boiling point associated with deoxygenation. The higher extent of deoxygenation encountered in this stage suggests that deoxygenation is greatly responsible in determining the boiling point distribution.

Boiling point distribution keeps on changing also in the second stage, where further cracking is achieved. From Table 5, it can be observed that the second stage enhances the production of very light distillates in the gasoline range, while fractions with boiling points higher than 320 °C decrease, particularly in the VGO range. For sewage sludge, the choice of the first stage temperature does not seem to influence the yields of fuel fractions in a significant fashion. In other words, the boiling point distribution seems to be a function of the highest temperature at which hydrotreating was carried out, as it is also testified by the overlapping curves in Figure 6b. This is not completely true for *Spirulina* biocrude, for which a higher temperature in the first stage results in higher yields of gasoline in the second, at the expense of heavy diesel and VGO fractions.

These results are directly correlated with the different heteroatoms removal efficiencies. For *Spirulina* biocrude, a significant part of denitrogenation is taking place in the second stage, and it is evident from Figure 4a that a higher temperature in the first stage results in a higher de-N in the final product. Cracking can be thus associated to the cleavage of nitrogen bonds (e.g. the peptide bond between two amino acids), leading to smaller compounds [50]. For sewage sludge biocrudes, final de-N is not substantially affected by the first stage temperature (Figure 4b) and cracking follows the same trend, as well.

TABLE 5

FIGURE 6

GC-MS analysis was applied to gain insight into the volatile fractions (boiling point of ca. 350 °C) of treated and untreated oil samples. Figure 7 shows the presence of different chemical families and their abundance on the basis of relative peak area for both raw biocrudes and hydrotreated products.

The volatile fractions of both biocrudes are mainly composed of oxygenated and nitrogenated compounds. For *Spirulina* biocrude there is a minor presence of *n*-paraffinic hydrocarbons (mostly due to *n*-$C_{17}$), while the majority of identified compounds are represented by O- and N,O-containing compounds, such as fatty acid amides. In sewage sludge biocrude, oxygenated compounds (mainly fatty acids) are the largest majority, although around 10% of the chromatogram area is explained by amides. Results are in good accordance with previous literature data [3,4,31].

After hydrotreating, a notable aspect is the abrupt change in the chemical composition of both biocrudes. Significant amounts of the heteroatom-containing compounds present in both biocrudes were converted into hydrocarbons, mostly paraffinic compounds (n-, iso-, and cyclo-paraffins). As it can be observed in Figure 7, *n*-paraffins represent a significant share of the compounds found after hydrotreating. This is especially true for sewage sludge (Figure 7b), where the relative share of *n*-paraffins after hydrotreating reached percentages of 70-80%. The presence of these straight-chain paraffins is directly connected to the nature of the compounds in the biocrudes. In this case, it is likely to be caused by the conversion of fatty amides and fatty acids, which is promptly achieved even at relatively mild reaction conditions.

After the second stage, the share of *n*-paraffins usually slightly decreases. This aspect seems to be caused by the generally increased portion of compounds detected by GC-MS after hydrotreating, due to the shift of the boiling point distribution to lower temperatures. Moreover, this speculation is also confirmed if compared with H/C and N/C atomic ratio (Figure 5). After the second stage H/C atomic ratio of both biocrudes is significantly increased, with a decrease in N/C atomic ratio. This means that the upgraded oil is more paraffinic due to the higher saturation of aromatic and heterocyclic compounds.

However, it is remarkable to observe that both iso- and cyclo-paraffins increase after the second stage. As far as cyclo-paraffins are concerned, their increase is especially remarkable in *Spirulina* biocrudes (Figure 7a), even though it is present in sewage sludge biocrude as well. Cyclo-paraffins are derived from the saturation of aromatic rings [10]. Iso-paraffins and olefins are probably derived from the cracking of terpenic fragments in the heavier fraction (larger molecules) of the biocrudes, although more research and powerful analytical tools (due to limitation of GC-MS for heavier oil fraction) are needed to prove this. Nevertheless, olefins detection in the upgraded product could be also due to inaccuracies in the spectra identification, as their mass spectra can be similar to those of other hydrocarbons.

In view of fuel properties, the increased share of iso- and cyclo-paraffins is highly beneficial, especially in view of the production middle distillates (jet-fuel and diesel). Indeed, these compounds can improve the so-called cold-flow properties (i.e. cloud point and pour point), significantly expanding the drop-in potential of these oils.

As far as nitrogen removal is concerned, it can be observed that nitrogen-containing compounds were not detected by GC-MS products after the second stage of hydrotreating. On the other hand, results from elemental analysis (Table 2) clearly show that nitrogen is still present in all samples. This corroborates the conclusion that the remaining nitrogen-containing compounds are actually found in the higher boiling point fractions, which are not enough volatile to be analyzed through GC-MS [19]. The absence of nitrogen-containing compounds from the volatile fraction of the upgraded oils, corresponding to range from which gasoline, jet-fuel and diesel are derived, represents a point in favor of their utilization for drop-in fuels production.

FIGURE 7

## 4. Conclusions

During the hydrotreatment of nitrogen-rich HTL biocrudes, the selection of optimized operating conditions is currently a challenge. Severe thermal conditions ($\geq$ 400 °C) are associated with effective hydrodenitrogenation. However, these severe conditions promote undesired polymerization or coking reactions, due to the presence of considerable amount of oxygen-containing reactive functional groups. Therefore, a more realistic pathway is followed, in which nitrogen-rich HTL biocrudes are first mildly hydroprocessed to a stabilized form, which can then be further upgraded with reduced coke formation. Results of the experiments suggest that oxygen-containing compounds in the HTL biocrude need to be removed first, in a stabilization stage. This subsequently allows a higher degree of denitrogenation and a reduction in coke formation.

The results from DSC provide important information regarding the thermal stability of *Spirulina* and sewage sludge biocrudes. In this study, it was found that *Spirulina* and sewage sludge biocrudes are thermally stable at 350 °C, but unstable at 400 °C. This information was then confirmed by observing the strong effect of temperature on coke production during the direct hydrotreatment at 350 °C and 400 °C. However, the mildly upgraded oil in a subsequent second stage at 400 °C did not contribute to overall coke production, but revealed to be extremely effective for nitrogen removal. Based on our findings, a hydrotreating process involving stabilization of biocrude at 350 °C and a subsequent second stage at 400 °C turned out to produce high quality bio-fuels with complete deoxygenation and a high degree of denitrogenation (around 92%). It also reduced coke yields when compared to direct hydroprocessing at 400 °C (i.e. 9.1% to 1.0% for *Spirulina* and 3.4% to 0.7% for sewage sludge biocrude). Two-stage processing can be therefore a viable strategy to deal with nitrogen-rich feedstocks in view of drop-in biofuel production.

**Acknowledgments**

This project has received funding from the European Union's Horizon 2020 research and innovation programme under grant agreement no. 764734 (HyFlexFuel).

**Table 6**

Summary of performed hydrotreating batch experiments. Each test was performed with an initial $H_2$ pressure of 8.0 MPa and a reaction time of 4 h.

| Experiment | No. of stage | Temperature (°C) |
|---|---|---|
| *Microalga Spirulina biocrude (AL)* | | |
| AL-310 | $1^{st}$ stage | 310 |
| AL-310-400 | $2^{nd}$ stage | 400 |
| AL-330 | $1^{st}$ stage | 330 |
| AL-330-400 | $2^{nd}$ stage | 400 |
| AL-350 | $1^{st}$ stage | 350 |
| AL-350-400 | $2^{nd}$ stage | 400 |
| AL-400 | $1^{st}$ stage | 400 |
| | | |
| *Sewage sludge biocrude (SS)* | | |
| SS-310 | $1^{st}$ stage | 310 |
| SS-310-400 | $2^{nd}$ stage | 400 |
| SS-330 | $1^{st}$ stage | 330 |
| SS-330-400 | $2^{nd}$ stage | 400 |
| SS-350 | $1^{st}$ stage | 350 |
| SS-350-400 | $2^{nd}$ stage | 400 |
| SS-400 | $1^{st}$ stage | 400 |

**Table 7**

Mass balance of the performed experiments. The results of all the experiments in 1$^{st}$ and 2$^{nd}$ stage were reported based on their inlet feed. The overall yields of 1$^{st}$ + 2$^{nd}$ stages were calculated based on the raw biocrudes (feed of the 1$^{st}$ stage).

| Experiment | Oil Yield (%) | Coke Yield (%) | Water Yield (%) | Gases Yield (%) | Total Balance (%) |
|---|---|---|---|---|---|
| **AL biocrude** | | | | | |
| AL-310 | 82.1±2.7 | 0.9±0.5 | 5.8±0.3 | 6.6±0.9 | 95.4±2.9 |
| AL-310-400 | 85.4±0.8 | 0.04±0.01 | 1.3±0.1 | 8.0±0.3 | 94.7±0.9 |
| Overall (1$^{st}$ + 2$^{nd}$ stage) | **70.1** | **1.0** | **6.9** | **13.2** | **91.3** |
| | | | | | |
| AL-330 | 80.2±4.9 | 1.6±0.1 | 6.1±0.2 | 6.8±0.6 | 94.6±4.9 |
| AL-330-400 | 85.5±1.1 | 0.03±0.01 | 0.9±0.1 | 10.3±0.3 | 96.7±1.1 |
| Overall (1$^{st}$ + 2$^{nd}$ stage) | **68.6** | **1.6** | **6.8** | **15.1** | **92.2** |
| | | | | | |
| AL-350 | 74.2±1.7 | 1.90±0.08 | 6.8±0.2 | 6.9±1.4 | 89.8±2.2 |
| AL-350-400 | 86.7±0.3 | 0.01±0.01 | 0.40±0.02 | 11.3±0.1 | 98.4±0.3 |
| Overall (1$^{st}$ + 2$^{nd}$ stage) | **64.3** | **1.9** | **7.1** | **15.3** | **88.7** |
| | | | | | |
| AL-400 | 63.5±0.5 | 9.1±0.2 | 7.4±0.3 | 15.0±1.2 | 95.1±1.3 |
| | | | | | |
| **SS biocrude** | | | | | |
| SS-310 | 81.1±0.8 | 0.6±0.1 | 3.1±0.6 | 9.1±1.6 | 93.9±1.9 |
| SS-310-400 | 82.9±0.4 | 0.05±0.01 | 7.6±0.4 | 7.9±0.7 | 99.1±0.9 |
| Overall (1$^{st}$ + 2$^{nd}$ stage) | **67.2** | **0.7** | **9.3** | **15.5** | **92.8** |
| | | | | | |
| SS-330 | 84.1±0.5 | 0.9±0.1 | 5.5±0.3 | 9.3±0.4 | 99.8±0.7 |
| SS-330-400 | 86.5±0.8 | 0.04±0.02 | 5.1±0.2 | 6.9±0.1 | 98.5±0.8 |
| Overall (1$^{st}$ + 2$^{nd}$ stage) | **72.7** | **1.0** | **9.8** | **15.1** | **98.7** |
| | | | | | |
| SS-350 | 73.7±1.9 | 1.51±0.02 | 10.9±0.2 | 11.5±1.8 | 97.6±2.6 |
| SS-350-400 | 87.5±0.8 | 0.014±0.008 | 0.70±0.01 | 5.0±0.1 | 93.2±0.8 |
| Overall (1$^{st}$ + 2$^{nd}$ stage) | **64.5** | **1.5** | **11.4** | **15.2** | **92.7** |
| | | | | | |
| SS-400 | 70.8±1.2 | 3.4±0.3 | 11.5±0.6 | 12.4±2.3 | 98.2±2.7 |

**Table 8**

Elemental composition, H/C atomic ratio and HHV of the biocrude and upgraded oil samples. Both de-O and de-N were calculated based on raw biocrude (feed of the 1$^{st}$ stage). Oxygen was calculated by difference.

| Sample | Elemental composition (%) | | | | de-O (%) | de-N (%) | H/C atomic ratio (-) | HHV (MJ kg$^{-1}$) |
| --- | --- | --- | --- | --- | --- | --- | --- | --- |
|  | C | H | N | O | | | | |
| AL biocrude | 75.0±0.3 | 10.4±0.1 | 7.7±0.1 | 6.9±0.1 | - | - | 1.66±0.02 | 37.6±0.2 |
| AL-310 | 81.3±0.3 | 11.6±0.0 | 6.0±0.1 | 1.1±0.4 | 84.6 | 22.5 | 1.72±0.01 | 41.9±0.2 |
| AL-310-400 | 84.5±0.1 | 12.9±0.0 | 2.4±0.1 | 0.0±0.1 | 100 | 68.6 | 1.83±0.00 | 44.7±0.3 |
| AL-330 | 81.8±0.6 | 11.8±0.1 | 4.9±0.4 | 1.5±0.7 | 78.3 | 36.5 | 1.73±0.03 | 42.2±0.3 |
| AL-330-400 | 84.7±0.0 | 13.4±0.0 | 1.8±0.0 | 0.0±0.1 | 100 | 76.4 | 1.89±0.00 | 45.3±0.3 |
| AL-350 | 83.6±0.3 | 12.9±0.1 | 3.5±0.1 | 0.0±0.3 | 100 | 54.3 | 1.85±0.01 | 44.3±0.2 |
| AL-350-400 | 84.8±0.2 | 14.6±0.0 | 0.6±0.0 | 0.0±0.2 | 100 | 92.5 | 2.07±0.00 | 46.8±0.2 |
| AL-400 | 83.6±0.1 | 12.1±0.2 | 4.4±0.4 | 0.0±0.4 | 100 | 43.4 | 1.74±0.03 | 43.3±0.3 |
| SS biocrude | 74.5±0.1 | 10.6±0.0 | 3.9±0.0 | 11.0±0.2 | - | - | 1.71±0.00 | 37.4±0.3 |
| SS-310 | 82.4±0.2 | 12.6±0.0 | 3.3±0.2 | 1.7±0.3 | 84.4 | 15.1 | 1.84±0.02 | 43.4±0.3 |
| SS-310-400 | 84.7±0.2 | 14.3±0.0 | 1.0±0.1 | 0.0±0.3 | 100 | 74.9 | 2.02±0.00 | 46.4±0.2 |
| SS-330 | 84.4±0.3 | 13.3±0.0 | 2.3±0.4 | 0.0±0.5 | 99.7 | 40.4 | 1.88±0.01 | 45.0±0.2 |
| SS-330-400 | 84.8±0.5 | 14.3±0.1 | 0.9±0.1 | 0.0±0.5 | 100 | 76.5 | 2.02±0.00 | 46.4±0.1 |
| SS-350 | 84.3±0.3 | 14.1±0.1 | 1.6±0.0 | 0.0±0.3 | 100 | 58.8 | 2.01±0.00 | 46.0±0.4 |
| SS-350-400 | 84.5±0.1 | 15.2±0.0 | 0.3±0.1 | 0.0±0.1 | 100 | 92.1 | 2.16±0.00 | 47.4±0.1 |
| SS-400 | 85.4±0.1 | 13.6±0.2 | 1.0±0.1 | 0.0±0.2 | 100 | 73.4 | 1.91±0.03 | 45.8±0.2 |

**Table 9**

Carbon balance of the performed experiments. Carbon yields (%) are reported with respect to carbon in the feed. The overall carbon balance of 1st + 2nd stages were calculated based on the raw biocrudes (feed of the 1st stage). Coke was assumed to be of pure carbon (C). C in the aqueous phase was not quantified.

| Experiment | Oil (% C) | Gas (% C) CO-$CO_2$ | Gas (% C) Hydrocarbons | Coke (% C) | Total (% C) |
|---|---|---|---|---|---|
| **AL biocrude** | | | | | |
| AL-310 | 89.02 | 0.02 | 0.96 | 0.92 | 90.9 |
| AL-310-400 | 88.74 | 0.02 | 1.55 | 0.04 | 90.4 |
| Overall (1st + 2nd stage) | **78.99** | **0.03** | **2.33** | **0.95** | **82.95** |
| | | | | | |
| AL-330 | 87.51 | 0.03 | 1.21 | 1.55 | 90.3 |
| AL-330-400 | 88.53 | 0.01 | 1.37 | 0.03 | 89.9 |
| Overall (1st + 2nd stage) | **77.47** | **0.04** | **2.42** | **1.57** | **81.50** |
| | | | | | |
| AL-350 | 82.68 | 0.01 | 1.78 | 1.90 | 86.4 |
| AL-350-400 | 87.96 | 0.00 | 0.88 | 0.01 | 88.8 |
| Overall (1st + 2nd stage) | **72.72** | **0.01** | **2.55** | **1.91** | **77.19** |
| | | | | | |
| AL-400 | 70.87 | 0.01 | 1.22 | 9.10 | 81.2 |
| | | | | | |
| **SS biocrude** | | | | | |
| SS-310 | 89.64 | 0.30 | 0.58 | 0.63 | 91.2 |
| SS-310-400 | 85.31 | 0.01 | 1.21 | 0.05 | 86.6 |
| Overall (1st + 2nd stage) | **76.47** | **0.31** | **0.76** | **0.67** | **78.21** |
| | | | | | |
| SS-330 | 95.27 | 0.14 | 0.93 | 0.92 | 97.3 |
| SS-330-400 | 86.90 | 0.02 | 1.33 | 0.04 | 88.3 |
| Overall (1st + 2nd stage) | **82.78** | **0.16** | **2.08** | **0.95** | **85.97** |
| | | | | | |
| SS-350 | 83.39 | 0.19 | 0.98 | 1.51 | 86.1 |
| SS-350-400 | 87.71 | 0.00 | 1.03 | 0.01 | 88.8 |
| Overall (1st + 2nd stage) | **73.14** | **0.19** | **1.88** | **1.52** | **76.73** |
| | | | | | |
| SS-400 | 81.06 | 0.17 | 1.02 | 3.40 | 85.7 |

**Table 10**

Mass fractions (%) of raw and upgraded oils in simulated distillation cuts (i.e. ASTM D-7169 [37]).

| Sample | Gasoline (< 193 °C) | Jet-fuel (193-271 °C) | Light diesel (272-321 °C) | Heavy diesel (321-425 °C) | VGO (425-564 °C) | Residue (> 564 °C) |
|---|---|---|---|---|---|---|
| AL biocrude | 4.1 | 10.9 | 9.9 | 28.7 | 13.9 | 32.5 |
| AL-310 | 6.9 | 17.6 | 14.7 | 17.4 | 16.2 | 27.2 |
| AL-310-400 | 23.0 | 22.5 | 22.9 | 10.6 | 3.4 | 17.6 |
| AL-330 | 8.5 | 18.4 | 17.8 | 16.6 | 14.9 | 23.8 |
| AL-330-400 | 26.2 | 23.2 | 22.1 | 10.4 | 3.3 | 14.7 |
| AL-350 | 17.2 | 19.6 | 24.7 | 15.1 | 9.5 | 13.9 |
| AL-350-400 | 29.6 | 20.7 | 23.1 | 8.3 | 2.8 | 15.5 |
| AL-400 | 22.8 | 23.2 | 22.8 | 12.0 | 5.0 | 14.2 |
| | | | | | | |
| SS biocrude | 1.4 | 3.1 | 4.4 | 38.0 | 11.8 | 41.3 |
| SS-310 | 3.6 | 13.2 | 25.0 | 16.7 | 14.6 | 26.9 |
| SS-310-400 | 14.8 | 21.4 | 36.3 | 13.7 | 6.2 | 7.6 |
| SS-330 | 5.7 | 15.6 | 31.4 | 15.6 | 13.1 | 18.6 |
| SS-330-400 | 14.9 | 21.5 | 36.1 | 13.4 | 5.5 | 8.6 |
| SS-350 | 9.8 | 15.3 | 36.3 | 16.0 | 11.5 | 11.0 |
| SS-350-400 | 16.8 | 19.5 | 36.3 | 13.1 | 5.6 | 8.7 |
| SS-400 | 12.9 | 17.3 | 34.3 | 14.3 | 7.3 | 13.9 |

**Figure 8**

Isothermal DSC curves of *Spirulina* (algae) and sewage sludge biocrude at 350 °C and 400 °C.

**Figure 9**

Picture of raw *Spirulina* biocrude (a), upgraded *Spirulina* biocrude at 350 °C in 1$^{st}$ stage (AL-350) (b), and further upgrading at 400 °C in 2$^{nd}$ stage (AL-350-400) (c).

**Figure 10**

Comparison between the oil yield and coke produced during two-stage hydrotreatment of *Spirulina* (a) and sewage sludge (b) biocrude. The graphs show the results of 1$^{st}$ stage (stage 1) and the accumulated results of both 1$^{st}$ and 2$^{nd}$ stage (stage 2), based on raw biocrudes.

**Figure 11**

Degree of deoxygenation (de-O), degree of denitrogenation (de-N) and hydrogen consumption during the hydrotreatment of *Spirulina* (a) and sewage sludge (b) biocrude. Stage 2 results report the overall heteroatoms removals (i.e. based on 1$^{st}$ stage feeds) and H$_2$ consumptions (i.e. the sum of H$_2$ consumptions occurred in both stages).

**Figure 12**

Modified van Krevelen diagram showing molar ratio of H/C as a function of N/C, during the two-stage hydrotreatment of *Spirulina* (a) and sewage sludge (b) biocrudes.

**Figure 13**

Mass fractions (%) of raw and upgraded oils in simulated distillation cuts (i.e. ASTM D-7169 [37]).

**Figure 14**

Chemical composition (relative peak area) in terms of families of compounds from the GC-MS analysis of treated and untreated oil samples from *Spirulina* (a) and sewage sludge (b) biocrudes.

**Figure 1**

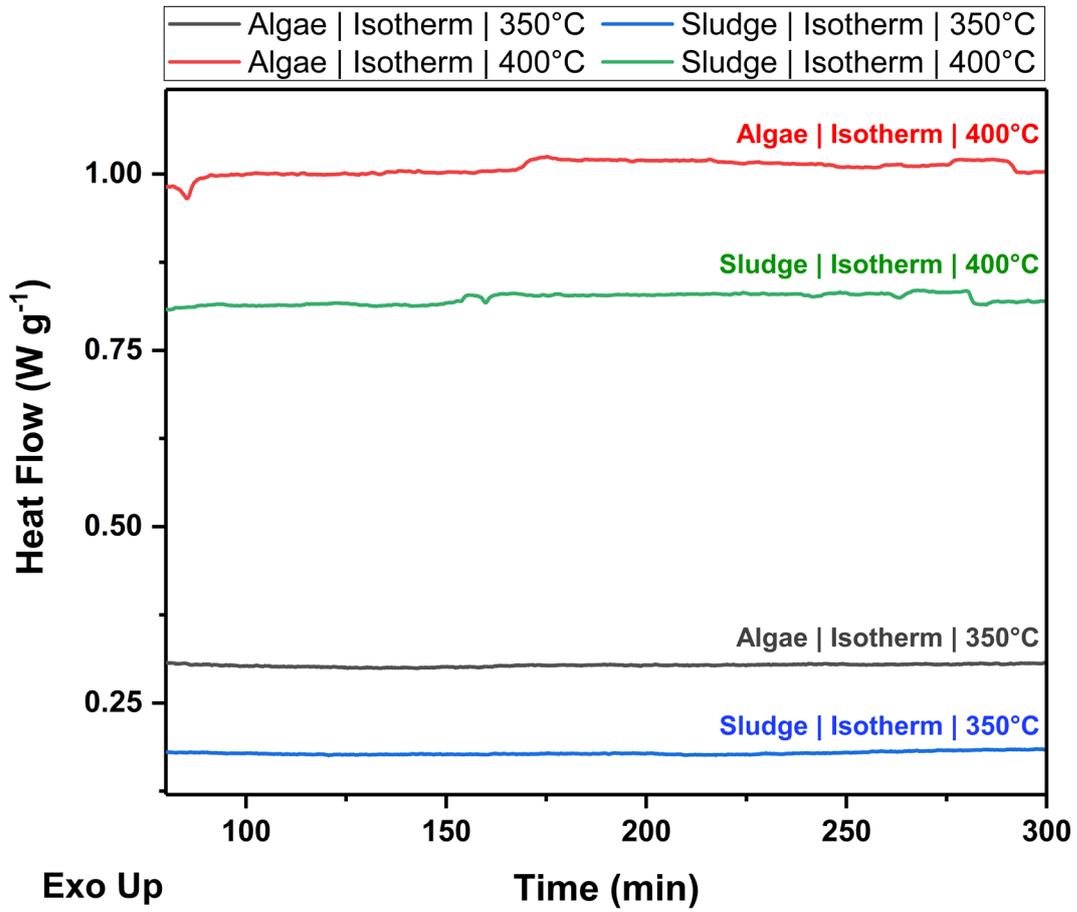

**Figure 2**

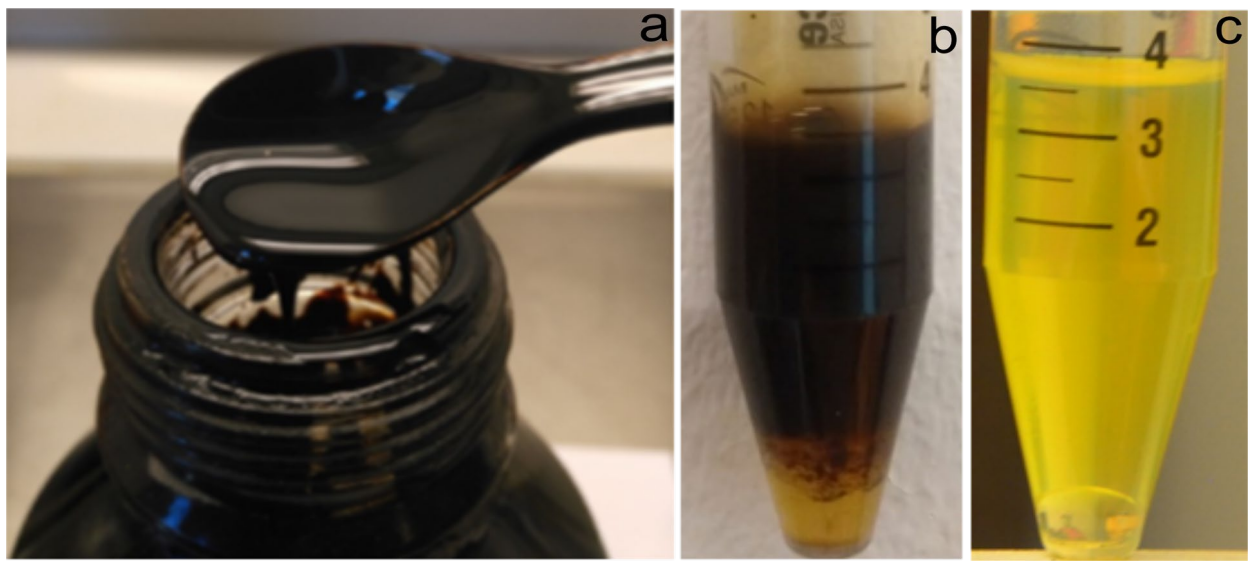

**Figure 3**

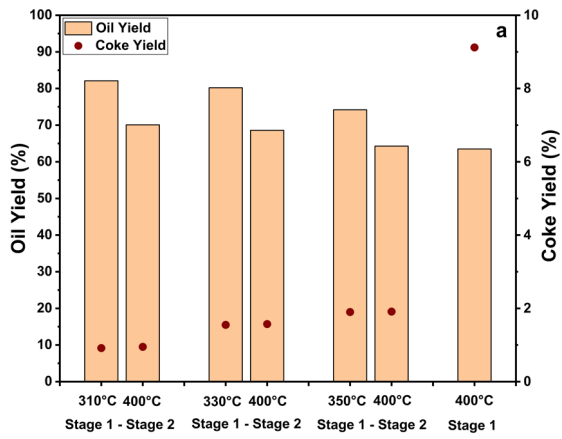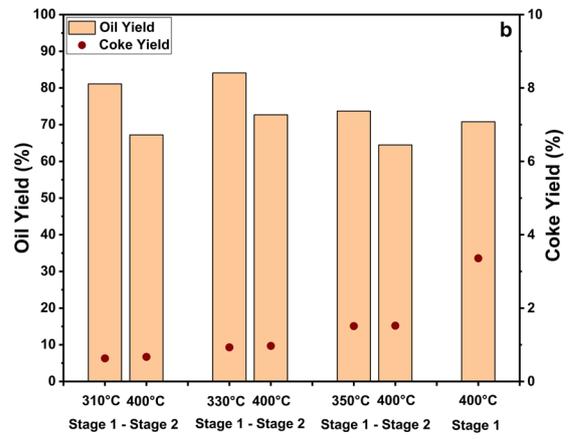

**Figure 4**

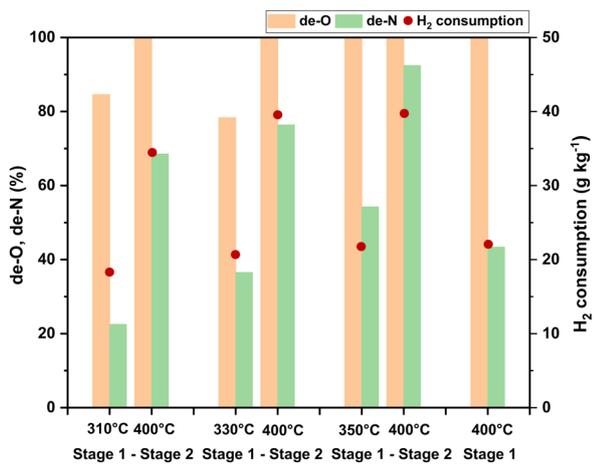
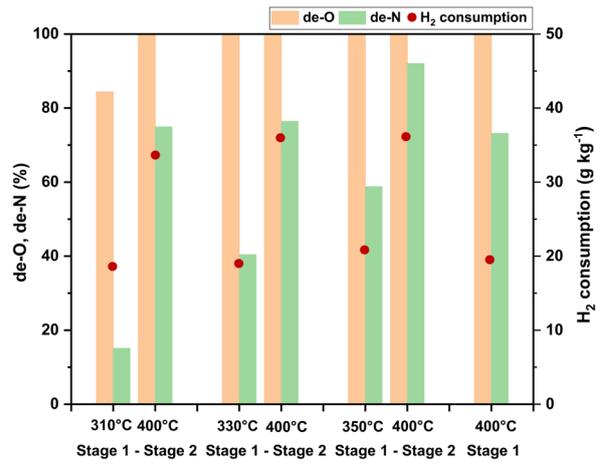

**Figure 5**

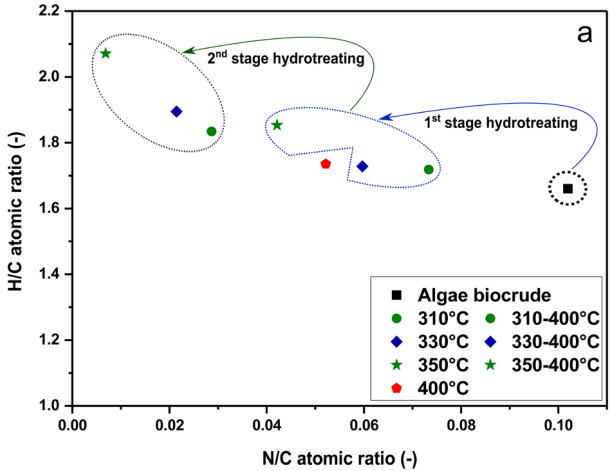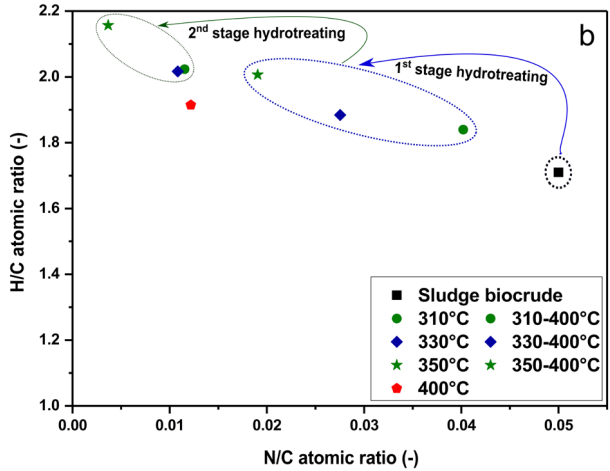

**Figure 6**

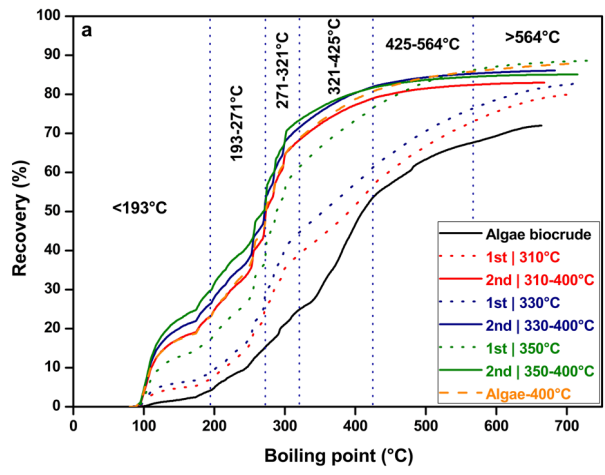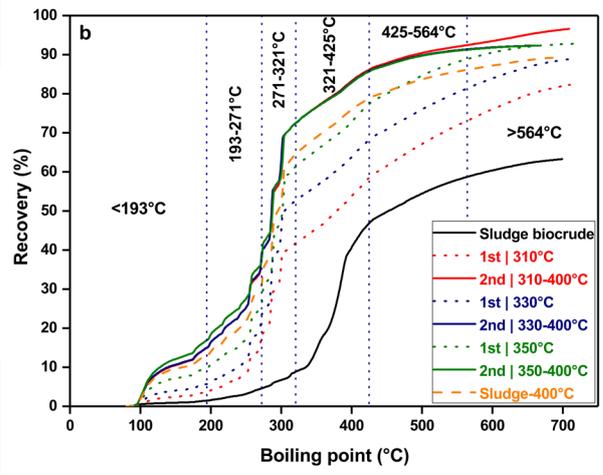

**Figure 7**

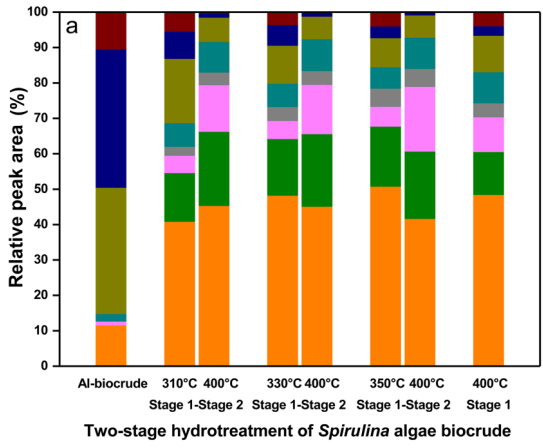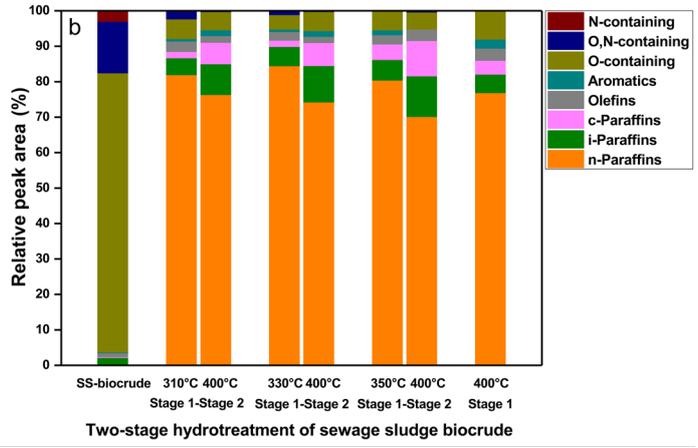